\documentstyle[preprint,aps]{revtex}

\begin{document}

\title{The nondeterministic Nagel-Schreckenberg traffic model with open 
boundary conditions}

\author{S. Cheybani$^{a,b}$ and J.Kert\'{e}sz$^{b,c}$ and M. 
Schreckenberg$^{a}$}

\address{$^a$
Theoretische Physik, Gerhard-Mercator Universit\"at, D-47048 Duisburg, Germany}
\address{$^b$
Department of Theoretical Physics, Technical University of Budapest, H-1111 
Budapest, Hungary}
\address{$^c$
Laboratory of Computational Engineering, Helsinki University of Technology, 
FIN-02150 Espoo, Finland}

\maketitle
\begin{abstract}
We study the phases of the Nagel-Schreckenberg traffic model with open 
boundary conditions as a function of the randomization probabilities p $>$ 0 
and the maximum velocity ${\mbox{v}}_{max}$ $>$ 1. Due to the existence of 
``buffer sites'' which enhance the free flow region, the behaviour is much 
richer than that of the related, parallel updated asymmetric exclusion process 
(ASEP, ${\mbox{v}}_{max}$ = 1). Such sites exist for 
${\mbox{v}}_{max}$ $\ge$ 3 and p $<$ ${\mbox{p}}_{c}$ where the phase diagram 
is qualitatively similar to the p = 0 case: there is a free flow and a 
jamming phase separated by a line of first order transitions. For 
p $>$ ${\mbox{p}}_{c}$ an additional maximum current phase separated by 
second order transitions occurs like for the ASEP. The density profile decays 
in the maximum current phase algebraically with an exponent $\gamma$ $\approx$ 
$\frac{2}{3}$ for all ${\mbox{v}}_{max}$ $\ge$ 2 indicating that these models 
belong to another universality class than the ASEP where $\gamma$ = 
$\frac{1}{2}$
\end{abstract}
\narrowtext

\section{Introduction}
\noindent
Asymmetric exclusion processes (ASEP) play an important role in 
non-equilibrium statistical mechanics. The one-dimensional ASEP is a 
lattice model which describes particles hopping in one direction with 
stochastic dynamics and hard core exclusion. It was first introduced in 
1968 to provide a qualitative understanding of the kinetics of the protein 
synthesis on RNA templates \cite{Don:Gi:Pi}. It turned out, however, that 
- despite its simplicity - there are numerous further applications of the 
ASEP on the field of interface growth, polymer dynamics, and traffic flow 
\cite{Schm:Zia}-\cite{Schr:Scha:Na:Ito}.\\
Unfortunately, as far as traffic is concerned, the ASEP yields rather 
unrealistic results, because essential phenomena like acceleration or 
slowing down cannot be reproduced in this model. 
As a consequence, Nagel and Schreckenberg developed an extension of the 
ASEP resulting in a one-dimensional probabilistic cellular automaton model 
\cite{Na:Schr}. According to the Nagel-Schreckenberg model 
the road consists of a single lane which is divided into L cells of equal size 
numbered by i = 1, 2, $\dots$, L and the time is also 
discrete. Each site can be either empty or occupied by a car with integer 
velocity v = 0, 1, $\dots$, ${\mbox{v}}_{max}$. All sites are simultaneously 
updated according to four successive steps: 
\begin{eqnarray}
&& \mbox{1. Acceleration: increase v by 1 if v $<$ 
         ${\mbox{v}}_{max}$.} 
\nonumber \nonumber \\
&& \mbox{2. Slowing down: decrease v to v = d if necessary 
         (d: number of empty cells in front of the car).} 
\nonumber \nonumber \\
&& \mbox{3. Randomization: decrease v by 1 with randomization 
         probability p if p $>$ 0.} 
\nonumber \nonumber \\
&& \mbox{4. Movement: move car v sites forward.} 
\nonumber 
\end{eqnarray}
Either ring (periodic boundary conditions) or 
open (open boundary conditions) geometry is considered. 
In the case of ring geometry cars move on a ring and the car density in the 
system keeps constant. Open systems, on the other hand, are characterized by 
the injection (extinction) rate $\alpha$ ($\beta$), that means, by the 
probability $\alpha$ ($\beta$) that a car moves into (out of) the system. \\
For the maximum velocity ${\mbox{v}}_{max}$ = 1 the model is identical 
with the ASEP with parallel update \cite{Ti:Er}-\cite{Ev:Ra:Spe}
which has been solved exactly with periodic boundary conditions 
\cite{Schr:Scha:Na:Ito} and recently also with open boundary conditions 
\cite{Ev:Ra:Spe}, \cite{deG:Nie}. 
In this special case three regimes (free flow, jamming, and maximum current) 
can be distinguished from each other. The transition from 
the free flow to the jamming phase at the $\alpha$ = $\beta$ - line for 
$\alpha$, $\beta$ $<$ 1-$\sqrt{\mbox{p}}$ is of first order. The transition 
from the free flow (jamming) to the maximum current phase is continuous and 
takes place at the injection (extinction) rate ${\alpha}_{c}$ (${\beta}_{c}$) 
with ${\alpha}_{c}$(p) = ${\beta}_{c}$(p) = 1-$\sqrt{\mbox{p}}$.\\
As it is common for traffic simulations cars are updated in parallel in the 
Nagel-Schreckenberg model, too, because this update scheme is the only one 
which models the formation of spontaneous jams occurring in real traffic 
\cite{Chowd:Sa:Scha}. Systems with parallel update are furthermore 
characterized by strong short-range correlations, and therefore, 
short-range correlation functions play an important role here 
\cite{Ra:Sa:Scha:Schr}, \cite{Chey1:Kert1:Schr1}.\\
Most of the work dealing with the Nagel-Schreckenberg model for 
${\mbox{v}}_{max}$ $>$ 1 impose periodic boundary conditions 
\cite{Vi:Sou}-\cite{Chey1:Kert1:Schr1}. 
Much attention has been paid to the question of the transition from the 
free flow to the jamming regime. According to the state-of-the-art this is 
a crossover rather than a sharp transition \cite{Lue:Schr:Us}-
\cite{Chowd:Kert:Na:Sa:Scha}. Systems with 
periodic boundary conditions are furthermore characterized by a trivial 
density profile $\rho$(i) =  $\rho$ with 1 $\le$ i $\le$ L due to 
translational invariance. In this context it should be mentioned that 
short-range correlation functions are well-suited for the description of 
the free flow - jamming transition 
\cite{Chey1:Kert1:Schr1}: The free flow regime is characterized 
by anticorrelations around a propagating peak, that is, in free flow cars 
are surrounded by empty space. At the critical density ${\rho}_{c}$ the 
anticorrelations are maximally developed, and for higher densities they 
vanish. Simultaneously, a jamming peak develops according to the fact that 
the  back car is strongly slowed down in a jam.\\
In the following, systems for maximum velocities 
${\mbox{v}}_{max}$ $\le$ 10 are investigated for the more realistic case of 
open boundaries. Boundary conditions are defined as in 
\cite{Chey2:Kert2:Schr2}: 
At site i = 0, that means out of the system a vehicle with the 
probability $\alpha$ and with the velocity v = ${\mbox{v}}_{max}$ is created. 
This car immediately moves according to the Nagel-Schreckenberg rules. 
If the velocity of the injected car on i = 0 is v = 0 (because site i = 1 is 
occupied by another car or because the front car is on site i = 2 and the 
injected car is slowed down by 1 due to randomization) then the injected car 
is deleted. 
At i = L+1 a ''block'' 
occurs with probability 1 - $\beta$ and causes a slowing down of the cars at 
the end of the system. Otherwise, with probability $\beta$, the cars simply 
move out of the system.
In \cite{Chey2:Kert2:Schr2} systems with open boundaries have been already 
analyzed for the randomization probability p = 0, i.e., by ignoring the 
randomization step. 
The most interesting feature of the deterministic Nagel-Schreckenberg model 
for maximum velocities ${\mbox{v}}_{max}$ $\ge$ 3 and open boundaries is the 
existence of so-called buffers: As a consequence of the parallel updating and 
the hindrance an injected car feels from the front car at the beginning of 
the system spaces larger than ${\mbox{v}}_{max}$  develop between two 
neighbouring cars for high injection rates. This can be easily demonstrated by 
considering i = 0 and the first sites of the system i = 1, $\cdots$, 5 for 
$\alpha$ = $\beta$ = 1 and ${\mbox{v}}_{max} \ge 3$. At an 
arbitrary time t = ${\mbox{t}}_{0}$ a car with velocity ${\mbox{v}}_{max}$ is 
injected on i = 0, i.e.
\begin{eqnarray}
injection: \hspace*{1cm}  {\mbox{v}}_{max}\mid .\,\ . \,\ 2 \,\ . \,\ . 
\nonumber 
\end{eqnarray}
After application of the NaSch rules on the system we have 
injected on i = 0, i.e.
\begin{eqnarray}
\hspace{-0.2cm} movement: \hspace*{1.6cm}  .  \mid . \,\ 2 \,\ . \,\ . \,\ . 
\nonumber 
\end{eqnarray}
Correspondingly, we get for t = ${\mbox{t}}_{0} + 1$
\begin{eqnarray}
&& injection: \hspace*{1cm}  {\mbox{v}}_{max}\mid .\,\ 2 \,\ . \,\ . \,\ . 
\nonumber \\ 
&& movement: \hspace*{1.5cm}  .  \mid 1 \,\ . \,\ . \,\ . \,\ 3 
\nonumber 
\end{eqnarray}
and for t = ${\mbox{t}}_{0} + 2$
\begin{eqnarray}
&& injection: \hspace*{1cm}  {\mbox{v}}_{max}\mid 1 \,\ . \,\ . \,\ . \,\ 3 
\nonumber \\ 
&& movement: \hspace*{1.4cm}  0  \mid . \,\ . \,\ 2 \,\ . \,\ . 
\nonumber 
\end{eqnarray}
As the car on site i = 0 cannot move it is deleted and the situation starts 
over in the next time step: 
\begin{eqnarray}
&& injection: \hspace*{1cm}  {\mbox{v}}_{max}\mid .\,\ . \,\ 2 \,\ . \,\ . 
\nonumber \\ 
&& movement: \hspace*{1.5cm}  .  \mid . \,\ 2 \,\ . \,\ . \,\ . 
\nonumber 
\end{eqnarray}
and so on. Obviously, one car is lost out of three injection possibilities. 
In the system, far from the boundaries, the distance between two neighbouring 
cars turns out to be alternately ${\mbox{d}}_{1} = {\mbox{v}}_{max}$ and 
${\mbox{d}}_{2} = 2({\mbox{v}}_{max}-1)$. 
That means, in addition to the 
expected ${\mbox{v}}_{max}$ empty sites larger gaps occur in the 
$\alpha$ $\rightarrow$ 1, $\beta$ $\rightarrow$ 1 limit. We call these 
additional sites ''buffers'' because they have a buffer effect at the end of 
the system: Due to these sites the development of jamming waves is suppressed 
even for $\beta$ $<$ 1. The transition from the free flow to the jamming 
phase is of first order and accompanied by the collapse of the buffers. The 
effect resulting from the buffers do not depend on the maximum velocity if 
${\mbox{v}}_{max}$ $\ge$ 5 (for ${\mbox{v}}_{max}$ = 3,4 the buffer effect 
is not so strong as the buffers are not completely developed for that case). 
\\
However, randomization is indispensable for the analysis of real traffic 
as it takes human behaviour into account: The behaviour of a car driver is 
not like that of a machine but rather contains unpredictable elements. 
In traffic, over-reactions when slowing down can be found as well as delays 
when accelerating, furthermore fluctuations when following a car 
(follow-the-leader situation) and so on.\\
Besides this motivation it is of interest to compare the general p $>$ 0, 
${\mbox{v}}_{max}$ $\ge$ 1 case with the previously investigated models 
(p $>$ 0, ${\mbox{v}}_{max}$ = 1 \cite{Ev:Ra:Spe}, \cite{deG:Nie} and 
p = 0, ${\mbox{v}}_{max}$ $\ge$ 1 \cite{Chey2:Kert2:Schr2}). 
The presented results were obtained by simulating a L = 1024 sample with 
at least 1000 runs with ${10}^{4}$ time steps each. 
In order to investigate the influence of randomization on the system we 
proceed similarly to \cite{Chey2:Kert2:Schr2}: Section II considers the 
behaviour of current and average occupation number in the middle of the 
system. Section III deals with density profiles, Section IV with short-range 
correlation functions. Finally, the results are summarized in Section V.

\section{Current and Average Occupation number at the middle of the system}
\noindent
The phase diagram for systems with probability p = 0.5 and
maximum velocity ${\mbox{v}}_{max}$ = 2 and ${\mbox{v}}_{max}$ $\ge$ 5 is 
shown in Figs 1a,b (the case ${\mbox{v}}_{max}$ = 3,4 is similar to the 
latter case). For ${\mbox{v}}_{max}$ = 2 the phase diagram 
(Fig 1a) is qualitatively the same as for 
the case ${\mbox{v}}_{max}$ = 1 \cite{Ev:Ra:Spe}, \cite{deG:Nie}: 
The free flow and the jamming regime are divided by a straight line and
for $\alpha$ $>$ 0.35 and $\beta$ $>$ 0.8 the system is in the 
maximum current phase. Obviously, the maximum current regime is 
smaller than for ${\mbox{v}}_{max}$ = 1 and there is no symmetry along 
the $\alpha$=$\beta$-line. 
For ${\mbox{v}}_{max}$ = 3 the maximum current regime is even 
smaller ( for $\alpha$ $>$ 0.35 and $\beta$ $>$ 0.85 ) and the 
free flow/jamming border shows a slight bending. These tendencies are even 
stronger developed for 
higher maximum velocities ${\mbox{v}}_{max}$ with a maximum current regime 
for for $\alpha$ $>$ 0.35, $\beta$ $>$ 0.87, and ${\mbox{v}}_{max}$ = 4
(for $\alpha$ $>$ 0.35 and $\beta$ $>$ 0.89, and ${\mbox{v}}_{max}$ $\ge$ 5, 
see Fig 1b).\\ 
Another interesting feature of the nondeterministic case is that the course 
of the free flow/jamming border is totally different from that for p = 0. 
In Fig 1c it can be clearly seen that this difference is a consequence of the 
vanishing buffer effect due to randomization. For randomization probabilities 
p $>$ ${\mbox{p}}_{c}$ (${\mbox{p}}_{c}$ = 0.1172 $\pm$ 0.008 for 
${\mbox{v}}_{max}$ = 5) there is no 
sign of the buffer effect any more, and a (rectangular) maximum current 
regime develops for $\alpha$ $>$ ${\alpha}_{Free Flow}$ and 
$\beta$ $>$ ${\beta}_{Jamming}$. \\
\\
In order to understand the nature of the transition between the phases 
we consider the average occupation number on the site 
i = $\frac{\mbox{L}}{\mbox{2}}$, $\rho$ (i = $\frac{\mbox{L}}{\mbox{2}}$), as 
proposed in \cite{Be:Cha:Ez}. Fig 2 shows 
$\rho$ (i = $\frac{\mbox{L}}{\mbox{2}}$) for ${\mbox{v}}_{max}$ = 5 
as a function of the injection and the extinction 
rates (the average occupation number on i = $\frac{\mbox{L}}{\mbox{2}}$ for 
any ${\mbox{v}}_{max}$ $>$ 1 behaves similarly). 
It turns out that phase transitions in systems with maximum velocity 
${\mbox{v}}_{max}$ $>$ 1 show the following features: 
The transition from free flow to jamming is of first order indicated by a 
jump in $\rho$ (i = $\frac{\mbox{L}}{\mbox{2}}$). At the 
maximum current/free flow and the maximum current/jamming transition there 
is a jump in the {\it derivative} of $\rho$ (i = $\frac{\mbox{L}}{\mbox{2}}$) 
what is a hint at a continuous phase transition. These conclusions will be 
confirmed in the next section which deals with the investigation of the 
corresponding density profiles. \\
\\
In the following we analyze the current due to the influence of the 
\begin{eqnarray}
&& \mbox{- boundaries} \nonumber \\
&& \mbox{- maximum velocity ${\mbox{v}}_{max}$} \nonumber \\
&& \mbox{- randomization probability p} \nonumber 
\end{eqnarray}
The best way to investigate the influence of the left boundary is to 
consider the case $\beta$ = 1 for p = 0.5 where cars simply move out of the 
system. In Fig 3a free flow and maximum current phase can be clearly 
distinguished from each other. For maximum velocities 
${\mbox{v}}_{max}$ $\ge$ 5 the curves are
nearly the same with a maximum at $\alpha$ $\approx$ 0.35 becoming 
stronger with increasing ${\mbox{v}}_{max}$ (the occurrence of the 
maximum will be explained below). The dependence of the current on 
$\alpha$ and $\beta$ for 
${\mbox{v}}_{max}$ = 2, 3, 4 is qualitatively the same as for 
${\mbox{v}}_{max}$ = 1.\\
For the investigation of the influence of the right boundary we consider 
the case $\alpha$ = 1 for p = 0.5 (Fig 3b). Here, the current for 
${\mbox{v}}_{max}$ $\ge$ 5 does not depend on the maximum velocity. 
Furthermore, it seems to increase monotonously with increasing $\beta$ 
also in the maximum current phase ($\beta$ $>$ 0.89). Investigations for 
system sizes L $\ge$ 4096, however, show that the latter observation is just 
a finite size effect and that the current for ${\mbox{v}}_{max}$ $\ge$ 5 and 
$\beta$ $>$ 0.89 is constant. 
Apart from this, the curves for the current do not change in an essential way 
with increasing system size L, and therefore it is sufficient to investigate 
systems with L = 1024 in the following. \\ 
From the observations so far we can conclude that - as for the deterministic 
case - the behaviour of the system only negligibly changes when maximum 
velocities ${\mbox{v}}_{max}$ $\ge$ 5 are considered. Therefore we 
restrict ourselves to the case ${\mbox{v}}_{max}$ = 5 in the following 
observations.
\\
\\
We will now investigate the influence of the randomization probability 
on the behaviour of the system. It can be seen from Figs 4a,b that the 
buffer effect observed for the deterministic case vanishes with increasing 
p: 
For $\beta$ = 1 in Fig 4a the maximum at $\alpha$ $\approx$ 0.81 resulting 
from the existence of the buffers moves to the left and becomes weaker 
and weaker (In Fig 4a  we make an exception and consider the current at 
${\mbox{v}}_{max}$ = 10 instead of ${\mbox{v}}_{max}$ = 5, because for 
the latter case this effect is nearly invisible). 
For the injection rate $\alpha$ = 1, on the other hand, the buffer effect 
vanishes as soon as randomization probabilities 
p $>$ ${\mbox{p}}_{c}$ are considered.
\\
To sum it up it can be said that 
as a consequence of the buffer effect, the course of the current 
in the maximum current phase deviates from the expected (constant) behaviour 
for $\beta$ = 1, p = 0.5, and maximum velocities ${\mbox{v}}_{max}$ $\ge$ 5 
showing a slight maximum at $\alpha$ $\approx$ 0.35. 
Besides, there are strong indications that a continuous transition from the 
free flow (jamming) to a maximum current phase develops with increasing 
randomization probability on the $\beta$ = 1 - line ($\alpha$ = 1 - line). 
More convincing arguments for the existence of a maximum current phase, 
however,   will be given in the following sections. 
\section{Density Profiles}
\noindent
Our observations so far consider the behaviour of the whole system and of the 
site i = $\frac{\mbox{L}}{\mbox{2}}$. For the analysis of what happens on 
the other sites it is useful to investigate the density profiles. 
Of special interest in this context is the question in how far the density 
profiles reflect the transition between the phases. For that purpose we 
consider the density profiles for $\beta$ = 1 ($\rightarrow$ transition from 
free flow to maximum current), for $\beta$ = 0.7 ($\rightarrow$ transition 
from free flow to jamming), and for $\alpha$ = 1 ($\rightarrow$ transition 
from maximum current to jamming). It turns out that -- as in the case of 
${\mbox{v}}_{max}$ = 1 -- the free flow (jamming) phase can be divided into 
the regime AI and AII (BI and BII). The following investigations are confined 
to the randomization probability p = 0.5, L = 1024 and ${\mbox{v}}_{max}$ = 5 
(for ${\mbox{v}}_{max}$ $\ge$ 5 and L $>$ 1024 the density profiles are 
qualitatively the same).\\ 
In Fig 5a the transition from free flow to maximum current for $\beta$ = 1 
can be clearly seen. The free flow regime is characterized by oscillations at 
the beginning of the system dying out for i $\gtrsim$ 100 (if p = 0.5 and 
${\mbox{v}}_{max}$ = 5) due to randomization and the density profile becomes 
constant. It can therefore be said that randomization blurs the influence of 
the left boundary. 
In the maximum current regime we do not have any oscillations at all. 
Instead, an analytic decrease of the density is observed becoming stronger 
with increasing $\alpha$. This phenomenon can be easily understood as cars 
hinder each other at the beginning of the system for high injection rates: 
The higher the injection rate the stronger the hindrance. As a consequence 
the density profiles in the maximum current phase do not depend at all on 
$\alpha$ in the middle and at the end of the system. 
At the beginning of the system, however, the density profiles decay 
as ${\mbox{i}}^{-\gamma}$ with $\gamma$ $\approx$ 0.66 what is valid for 
{\it all} ${\mbox{v}}_{max}$ $>$ 1. We conjecture that 
$\gamma$ = $\frac{2}{3}$ because the exponent converges to this value with 
increasing system sizes. 
That means that the cases ${\mbox{v}}_{max}$ = 1 
and ${\mbox{v}}_{max}$ $>$ 1 belong to different universality classes since 
the corresponding density profiles for the ASEP with parallel update decay 
as ${\mbox{i}}^{-\gamma}$ with $\gamma$ = $\frac{1}{2}$ at the beginning of 
the system \cite{Ev:Ra:Spe}, \cite{deG:Nie}. \\ 
The free flow/maximum current transition is nicely reflected 
by the density at the end of the system: $\rho$(i=L) is proportional to the 
injection rate when the cars move freely up to ${\alpha}_{c}$ = 0.35 and 
becomes constant in the maximum current regime. 
\\
The situation is different when density profiles at the free flow - jamming 
border for $\beta$ = 0.7 are considered (Fig 5b). It can be easily seen that 
the transition is of first order as the density profile at 
${\alpha}_{c}$ = 0.278 is linear which is a typical feature of a first-order 
phase transition (see \cite{Chowd:Sa:Scha} and references therein).
\\
The density profiles for the injection rate $\alpha$ = 1 are shown in Fig 5c. 
The course of the curves for $\beta$ $<$ 0.89 ($\beta$ $>$ 0.89) is typical 
for the maximum current (BII jamming) phase with an algebraic (exponential) 
decay at the beginning of the system due to the hindrance 
already described for $\beta$ = 1. The decay becomes weaker with decreasing 
$\beta$ and finally vanishes as the repercussion resulting from the blockage 
at i = L+1 increasingly superimposes the hindrance effect at the beginning of 
the system. 
For extinction rates 0.6 $\lesssim$ $\beta$ $\lesssim$ 0.9 there is a slight 
increase at the end of the system which indicates a hindrance due to the 
blockage. It just remains a border effect, however, and is of no relevance 
for the considerations in this article.
\section{Correlation Functions}
\noindent
In this chapter we consider short-range correlation functions 
\begin{eqnarray}
\mbox{C(i,t) = $<\eta \mbox{(i',t')} \,\ \eta 
\mbox{(i+i',t+t')}>_{\mbox{i',t'}}$ - $<\eta \mbox{(i',t')}>_{\mbox{i',t'}}$}
^{2} \nonumber  
\end{eqnarray}
where
\begin{eqnarray}
&&\eta\mbox{(i',t') = 1} \,\ \,\ \mbox{  if site i' is occupied at time t '}
\nonumber\\
&&\eta \mbox{(i',t') = 0} \,\ \,\ \mbox{  else} \nonumber
\end{eqnarray}
$<...>_{\mbox{i',t'}}$ describes the spatial and temporal 
average over all L sites i' and over times t' taken from our simulation 
of the steady state.\\
The correlation functions are measured in the middle of the system where 
the influence of the boundaries is minimal. We do not only investigate the 
cases $\beta$ = 1 (influence of the left boundary, Fig 6a) and $\alpha$ = 1 
(influence of the right boundary, Fig 6b), but also $\beta$ = 1-$\alpha$ 
(Fig 6c). 
For the latter case there are similar conditions at both boundaries and the 
system can be compared at best with the corresponding system with periodic 
boundary conditions. Therefore it is no surprise that the correlation 
functions in Fig 6c are qualitatively the same as those for periodic 
boundary conditions \cite{Chey1:Kert1:Schr1}.\\
What is interesting, however, is that a classification into free flow, 
maximum current, and jamming cannot be done when short-range correlation 
functions are considered. Instead, due to Figs 6a-c three regimes can be 
distinguished from each other:
\begin{eqnarray} 
\mbox{{\bf (a)}} &&\mbox{{\bf Free Flow:} The free flow is characterized by 
anticorrelations around a } 
\nonumber \\ 
&&\mbox{propagating peak at i = ${\mbox{v}}_{max}$(t-1) with a shoulder at 
i = ${\mbox{v}}_{max}$ t, that is,} \nonumber \\
&&\mbox{ in free flow moving cars are surrounded by empty space}
\nonumber \\
\mbox{{\bf (b)}} &&\mbox{{\bf Coexistence Regime (JI + Maximum current):} 
The coexistence of } 
\nonumber \\  
&&\mbox{free flow and jamming manifests itself in the double peak structure of the } 
\nonumber \\ 
&&\mbox{correlation function. The jamming causes a maximum at i = -1 according } 
\nonumber \\
&&\mbox{to the hindrance the back car feels in the jam. } 
\nonumber \\   
\mbox{{\bf (c)}} &&\mbox{{\bf Jamming II (JII, ''Superjamming''):} 
The propagating peak disappears } \nonumber \\
&&\mbox{as a consequence of the fact that in free flow moving cars do not 
exist any } 
\nonumber \\
&&\mbox{longer.}
\nonumber
\end{eqnarray}
As it is obvious from the previous sections the transition from JI to JII is 
not a phase transition and does not change the behaviour of the system 
in an essential way.\\
In correspondence with \cite{Chey1:Kert1:Schr1} the critical injection 
(extinction) rate ${\alpha}_{c2}$ (${\beta}_{c2}$) for the JI-JII transition 
is defined by the vanishing of the propagating peak and takes place at 
${\beta}_{c2}$ $\approx$ 0.65 (see also Fig 1b). 
Unfortunately, an exact value for the Free Flow-Jamming transition can 
neither be given. As a consequence of the randomized oscillations in the 
density at the beginning of the system it is not possible to determine the 
critical injection (extinction) rate at which the influence of the right 
boundary reaches the left boundary. This is a significant difference to the 
deterministic case \cite{Chey2:Kert2:Schr2} where the oscillations of the 
free flow phase form a well-defined pattern due to the lack of randomization.
\\
The transition from Jamming I to Jamming II takes place at $\beta$ $\approx$ 
0.65 for {\it all} ${\mbox{v}}_{max}$ $>$ 1. In other words: When the maximum 
velocity is varied Free Flow, JII and coexistence phase (JI+maximum current) 
keep constant and only the ratio between maximum current phase and JI changes 
(Figs 1a,b).

\section{Conclusions}
\noindent
The nondeterministic Nagel-Schreckenberg model depends on the randomization, 
the maximum velocity, and the boundary conditions: 
\\
The buffer effect observed for the deterministic case p = 0 and 
${\mbox{v}}_{max}$ $\ge$ 3 is strongly weakened with increasing randomization 
probability. For ${\mbox{v}}_{max}$ = 2 there are no buffers and therefore 
the corresponding phase diagram is similar to the case 
${\mbox{v}}_{max}$ = 1 for all p values. 
For ${\mbox{v}}_{max}$ $\ge$ 3 and p $>$ ${\mbox{p}}_{c}$ 
(${\mbox{p}}_{c}$ = 0.1172 $\pm$ 0.008 for ${\mbox{v}}_{max}$ = 5) the buffer 
effect completely vanishes since the development of jamming waves is no longer 
suppressed. 
As a consequence, a maximum current phase occurs for p $>$ ${\mbox{p}}_{c}$ 
and the free flow (jamming) phase can be divided into two regimes AI and AII 
(BI and BII) similarly to the case of ${\mbox{v}}_{max}$ = 1,2. 
Another analogy to the case ${\mbox{v}}_{max}$ = 1 is that the 
free flow/jamming (free flow/maximum current and jamming/maximum current) 
transition is of first (second) order. \\ 
There are, however, essential differences between systems with 
${\mbox{v}}_{max}$ = 1 and ${\mbox{v}}_{max}$ $>$ 1: 
In the maximum current phase the density profiles decay algebraically with an 
exponent $\gamma$ = $\frac{2}{3}$ for ${\mbox{v}}_{max}$ $\ge$ 2 whereas 
$\gamma$ = $\frac{1}{2}$ was obtained in the ASEP. This indicates that 
systems with ${\mbox{v}}_{max}$ $>$ 1 and ${\mbox{v}}_{max}$ = 1 belong to 
different universality classes. Another difference to the ASEP is the 
existence of oscillations at the beginning of the system due to the repulsion 
of the cars. 
\\
The comparison of systems with periodic and with open boundary conditions 
suggests that there are mainly three differences. First of all, the transition 
from free flow to jamming for systems with open boundaries is sharp 
and there is no maximum current phase in the case of periodic boundary 
conditions. Moreover, the dependence on the maximum velocity is more complex 
for systems with open boundary conditions due to the occurrence of the buffers.
However, there are common features, too:
Measurements of the short-range correlation function show that -- as for 
corresponding systems with periodic boundary conditions -- three regimes 
can be distinguished from each other: (a) Free Flow: cars do not hinder each 
other (b) Maximum Current and Jamming I: coexistence of freely moving and 
jammed cars (c) Jamming II: cars are jammed.
\section{Acknowledgments}
This work was supported by the State of North Rhine-Westphalia and by 
OTKA(T029985).

\newpage

\section*{FIGURE CAPTIONS}

\noindent
{\bf Fig 1a:}\\
Phase diagram for ${\mbox{v}}_{max}$ = 2 
(p = 0.5, continuous line: first-order or continuous phase transition, 
dotted line: border between AI/AII and BI/BII, broken line: border between 
the JI and the JII regime). Although there is no symmetry along the 
$\alpha$ = $\beta$ - line the phase diagram shows strong similarities to the 
${\mbox{v}}_{max}$ = 1 case.
\\
\\
\noindent 
{\bf Fig 1b:}\\
Phase diagram for ${\mbox{v}}_{max}$ $\ge$ 5 
(p = 0.5, continuous line: first-order or continuous phase transition, 
dotted line: border between AI/AII and BI/BII, broken line: border between 
the JI and the JII regime). Phase diagrams for ${\mbox{v}}_{max}$ = 3,4 
are qualitatively the same. Due to the buffer effect the border between the 
free flow and the jamming phase shows a slight bending.
\\
\\
\noindent 
{\bf Fig 1c:}\\
Phase diagram for ${\mbox{v}}_{max}$ = 5 and p = 0, 0.1172, 0.25, and 0.5. 
For randomization probabilities p $>$ ${\mbox{p}}_{c}$ with 
${\mbox{p}}_{c}$ = 0.1172 $\pm$ 0.008 the buffer 
effect vanishes and the free flow regime becomes smaller with increasing 
randomization probability. The borders of the maximum current regime for 
$\alpha$ $>$ 0.6, $\beta$ $>$ 0.92, and p = 0.25 ($\alpha$ $>$ 0.35, 
$\beta$ $>$ 0.89, and p = 0.5) are represented by broken lines. 
The phase diagram for maximum velocities ${\mbox{v}}_{max}$ $>$ 5 is very 
similar. 
\\
\\
\noindent
{\bf Fig 2:}\\
Average density in the middle of the system for ${\mbox{v}}_{max}$ = 5 
(p = 0.5). The first-order phase transition from freely moving to jammed 
traffic can be clearly seen. Moreover, there is a jump in the derivative 
of $\rho$(i,$\frac{\mbox{L}}{\mbox{2}}$) at the maximum current/free flow 
and the maximum current/jamming border what is a hint at a continuous phase 
transition.
\\
\\
{\bf Fig 3a:}\\
Current q for ${\mbox{v}}_{max}$ = 2, 3, $\cdots$, 10 and $\beta$ = 1 
(p = 0.5). At $\alpha$ $\approx$ 0.35 there is a slight maximum for 
${\mbox{v}}_{max}$ $\ge$ 5 which is explained in Fig 4a.
\\
\\
{\bf Fig 3b:}\\
Current q for ${\mbox{v}}_{max}$ = 2, 3, $\cdots$, 10 and $\alpha$ = 1 
(p = 0.5). For ${\mbox{v}}_{max}$ $\ge$ 5 the curves are identical.
\\
\\
{\bf Fig 4a:}\\
Current q for p = 0, 0.125, $\cdots$, 0.875 and $\beta$ = 1 
(${\mbox{v}}_{max}$ = 10). The maximum at $\alpha$ $\approx$ 0.81 for p = 0 
moves towards smaller injection rates with increasing randomization 
probabilities and for p $>$ 0.125 a continuous phase transition is observed.
\\
\\
{\bf Fig 4b:}\\
Current q for p = 0, 0.125, $\cdots$, 0.875 and $\alpha$ = 1 
(${\mbox{v}}_{max}$ = 5). For p $>$ 0, the buffer effect observed for the 
deterministic case vanishes and the first-order phase transition goes over 
into a continuous phase transition.
\\
\\
{\bf Fig 5a:}\\
Density profiles for $\beta$ = 1 (p = 0.5, ${\mbox{v}}_{max}$ = 5). 
A typical feature of density profiles in the free flow regime are 
oscillations resulting from the hindrance the cars feel at the beginning of 
the system from each other. These oscillations die out for higher system 
sites due to randomization. The curves decay algebraically in the maximum 
current regime as it is already known from the ASEP.
\\
\\
{\bf Fig 5b:}\\
Density profiles for $\beta$ = 0.7 (${\mbox{v}}_{max}$ = 5, p = 0.5). 
The phase transition at $\alpha$ = 0.278 is of first order characterized 
by a linear density profile at the critical injection rate. The curve for 
$\alpha$ = 0.4 ($\alpha$ = 0.3 or $\alpha$ = 0.29) is typical for a density 
profile in the BII (BI) regime.
\\
\\
{\bf Fig 5c:}\\
Density profiles for $\alpha$ = 1 (p = 0.5, ${\mbox{v}}_{max}$ = 5). 
The results from the ${\mbox{v}}_{max}$ = 1 case are recovered: In the 
maximum current regime ($\beta$ $>$ 0.89) the density profiles decay 
algebraically and in the BII-jamming regime they are described by an enhanced 
exponential function.
\\
\\
{\bf Fig 6a:}\\
Correlation functions for $\beta$ = 1 (p = 0.5, ${\mbox{v}}_{max}$ = 5). 
The free flow regime is characterized by a propagating peak with 
anticorrelations around it. In the coexistence regime both the jamming and 
the propagating peak can be observed.
\\
\\
{\bf Fig 6b:}\\
Correlation functions for $\alpha$ = 1 (p = 0.5, ${\mbox{v}}_{max}$ = 5). 
The curves in the maximum current and in the JI-jamming regime behave 
similarly. At $\beta$ $\approx$ 0.75 the propagating peak vanishes and 
the system is in the JII-Jamming ( = ''Superjamming'') regime.
\\
\\
{\bf Fig 6c:}\\
Correlation functions for $\beta$ = 1-$\alpha$ (p = 0.5, 
${\mbox{v}}_{max}$ = 5). As for the injection rate  $\alpha$ = 1, 
the propagating peak vanishes at $\beta$ = 1-$\alpha$ $\approx$ 0.75. 
Moreover, there is a striking similarity to corresponding correlation 
functions in the case of periodic boundary conditions.


\begin{thebibliography}{4}
\bibitem {Don:Gi:Pi}
T J MacDonald, J H Gibbs, and A C Pipkin, 
{\it Biopolymers} {\bf 6}, 1 (1968)
%
\bibitem {Schm:Zia}
B Schmittmann and R K P Zia, 
{''Statistical Mechanics of Driven Diffusive Systems''} 
(Vol 17 of Domb and Lebowitz series), Academic Press (1995)
%
\bibitem {Hal:Zha}
T Halpin-Healy and Y C Zhang, 
{\it Phys Rep} {\bf 254}, 214 (1995)
%
\bibitem{Lee:Koo}
J M J van Leeuwen and A Kooiman, 
{\it Physica A} {\bf 184}, 79 (1992)
%
\bibitem{Na:Schr}
K Nagel and M Schreckenberg, 
{\it J Phys I} {\bf 2}, 2221 (1992)
%
\bibitem{Schr:Scha:Na:Ito}
M Schreckenberg, A Schadschneider, K Nagel, and N Ito, 
{\it Phys Rev E} {\bf 51}, 2939 (1995)
%
\bibitem{Ti:Er}
L G Tilstra and M H Ernst, 
{\it J Phys A} {\bf 31}, 5033 (1998)
%
\bibitem{Be:Cha:Ez}
A Benyoussef, H Chakib, and H Ez-Zahraouy, 
{\it Eur Phys J B} {\bf 8}, 275 (1999)
%
\bibitem {Ev:Ra:Spe}
M R Evans, N Rajewsky, and E R Speer, 
{\it J Stat Phys} {\bf 95}, 45 (1999) 
%
\bibitem {deG:Nie}
J de Gier and B Nienhuis, 
{\it Phys Rev E} {\bf 59}, 4899 (1999)
%
\bibitem {Chowd:Sa:Scha}
D Chowdhury, L Santen, and A Schadschneider, 
{\it cond-math/9910173}
%
\bibitem{Ra:Sa:Scha:Schr}
N Rajewsky, L Santen, A Schadschneider, and M Schreckenberg, 
{\it J Stat Phys} {\bf 92}, 151 (1998)
%
\bibitem{Vi:Sou}
L C Q Vilar and A M C de Souza, 
{\it Physica A} {\bf 211}, 84 (1994)
%
\bibitem{Csa:Kert}
G Cs\'anyi and J Kert\'esz, 
{\it J Phys A} {\bf 28}, L427 (1995) 
%
\bibitem{Ro:Lue:Us}
L Roters, S L\"ubeck, and K D Usadel, 
{\it Phys Rev E} {\bf 59}, 2672 (1999)
%
\bibitem{Lue:Schr:Us}
S L\"ubeck,M Schreckenberg, and K D Usadel, 
{\it Phys Rev E} {\bf 57}, 1171 (1998) 
%
\bibitem{Scha}
A Schadschneider, 
{\it Eur Phys J B} {\bf 10}, 573 (1999) 
%
\bibitem{Scha1:Schr1}
A Schadschneider and M Schreckenberg, 
{\it J Phys A} {\bf 26}, L679 (1993) 
%
\bibitem{Scha2:Schr2}
A Schadschneider and M Schreckenberg, 
{\it J Phys A} {\bf 30}, L69 (1997) 
%
\bibitem{Scha3:Schr3}
A Schadschneider and M Schreckenberg, 
{\it J Phys A} {\bf 31}, L225 (1998) 
%
\bibitem {Eis:Sa:Scha:Schr}
B Eisenbl\"atter, L Santen, A Schadschneider, and M Schreckenberg, 
{\it Phys Rev E} {\bf 57}, 1309 (1998)
%
\bibitem {Sas:Kert}
M Sasv\'ari and J Kert\'esz, 
{\it Phys Rev E} {\bf 56}, 4104 (1997)
%
\bibitem {Sa:Scha}
L Santen and A Schadschneider, 
{\it cond-math/9711261}
%
\bibitem {Chey1:Kert1:Schr1}
S Cheybani, J Kert\'esz, and M Schreckenberg, 
{\it J Phys A} {\bf 31}, 9787 (1998)
%
\bibitem {Chowd:Kert:Na:Sa:Scha}
D Chowdhury, J Kert\'esz, K Nagel, L Santen, and A Schadschneider,  
{\it Phys Rev E} {\bf 61}, 3270 (2000)
%
\bibitem {Chey2:Kert2:Schr2}
S Cheybani, J Kert\'esz, and M Schreckenberg, 
{\it to be published}
%
\end{thebibliography}
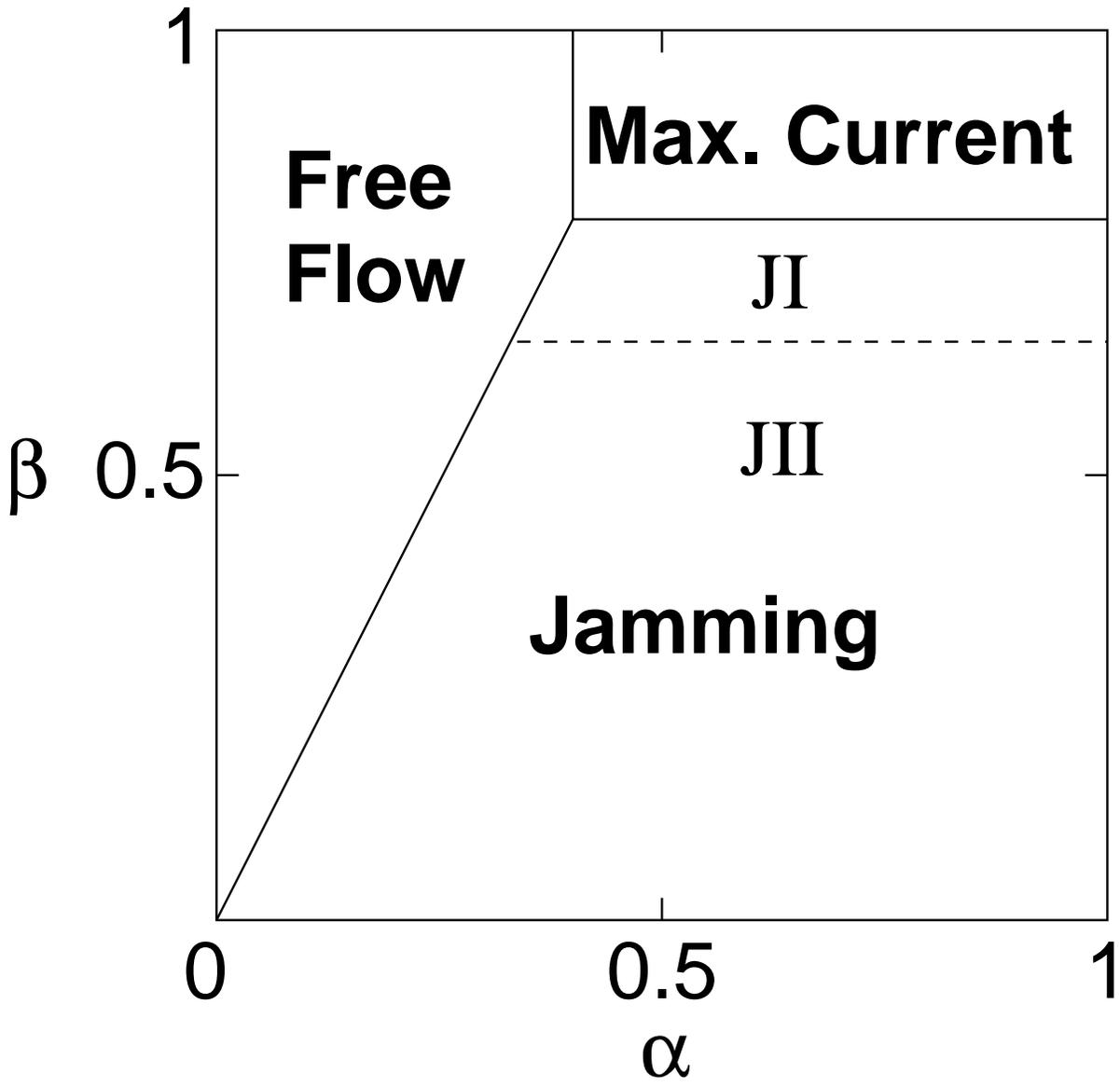
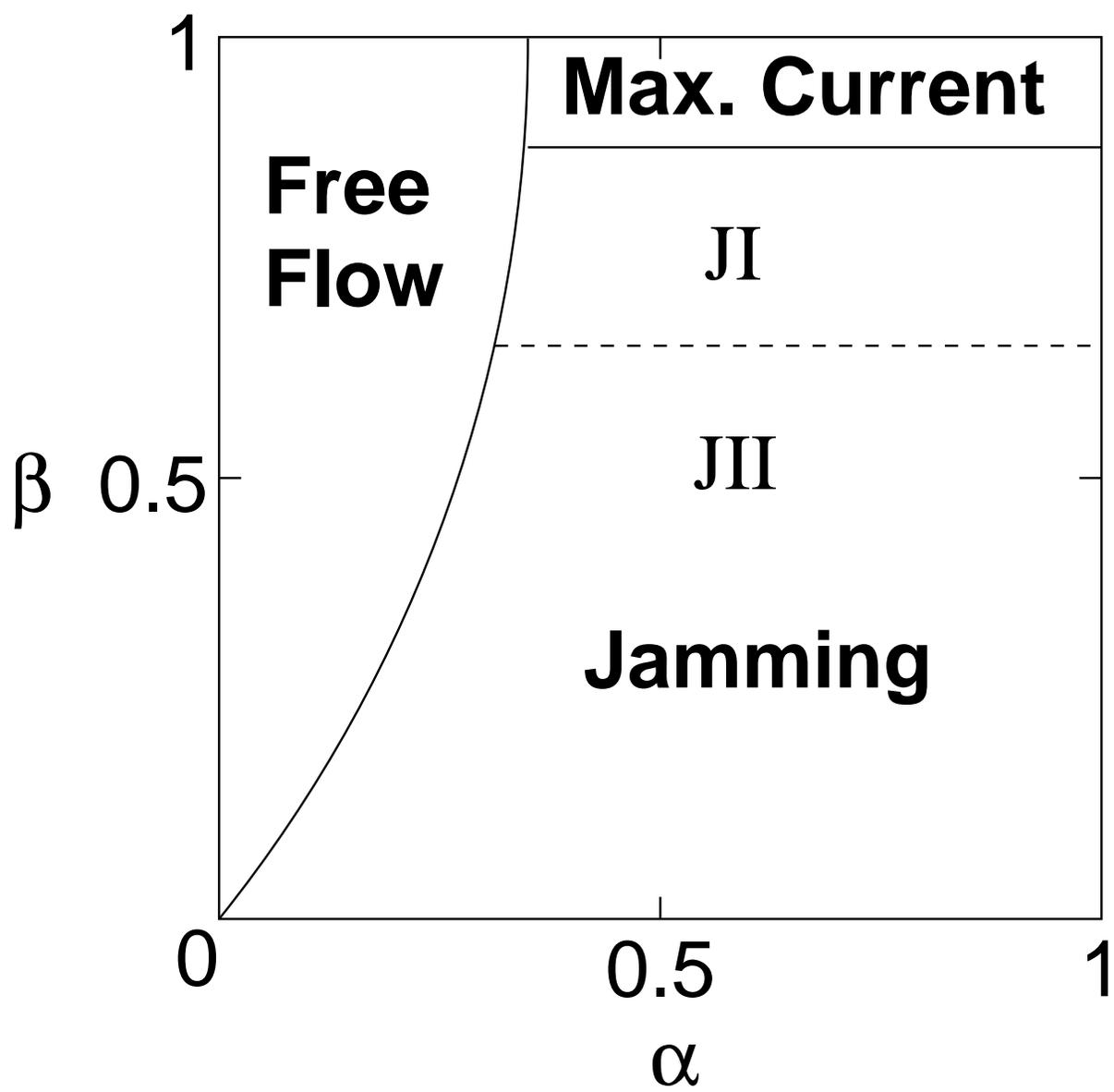
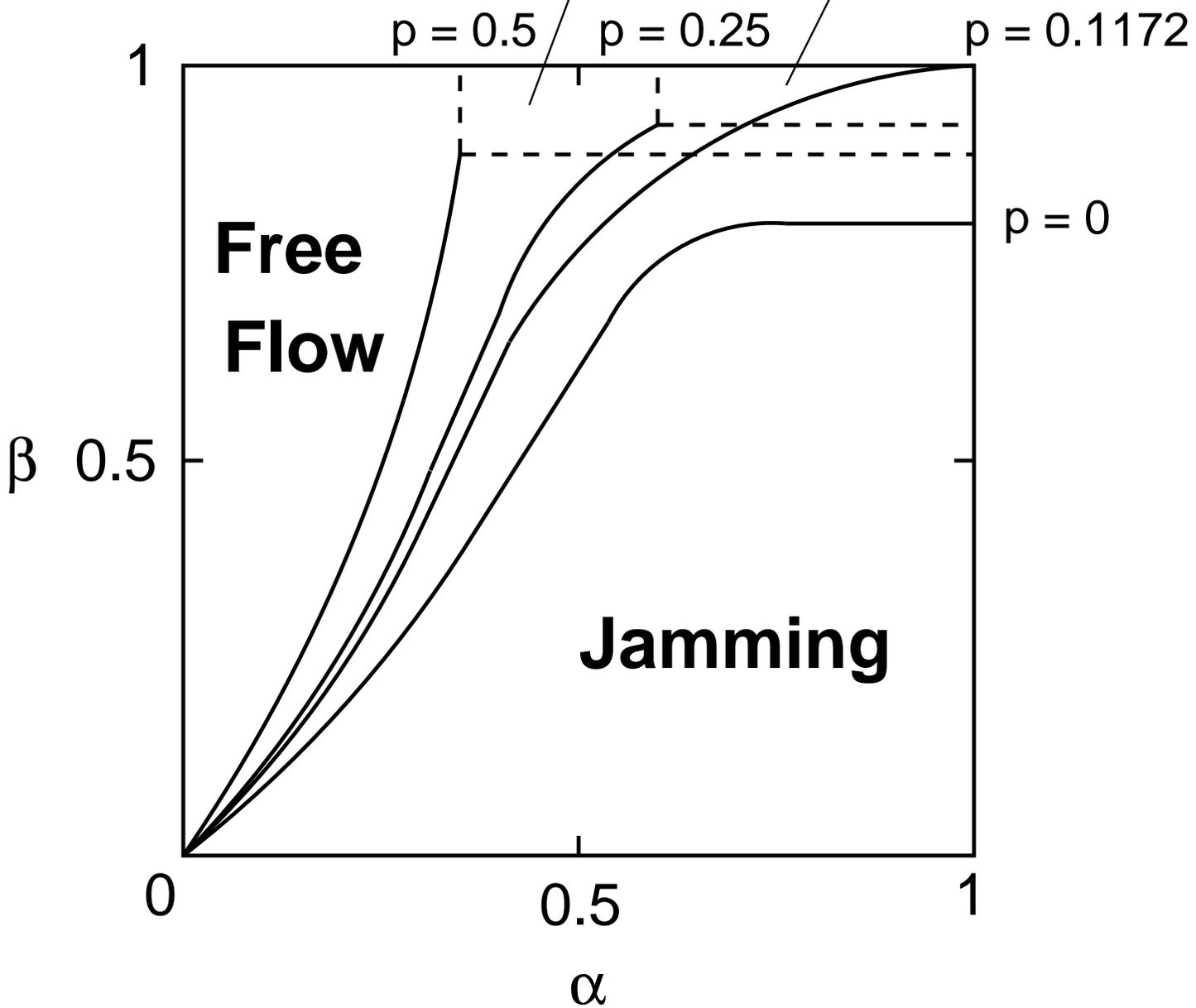
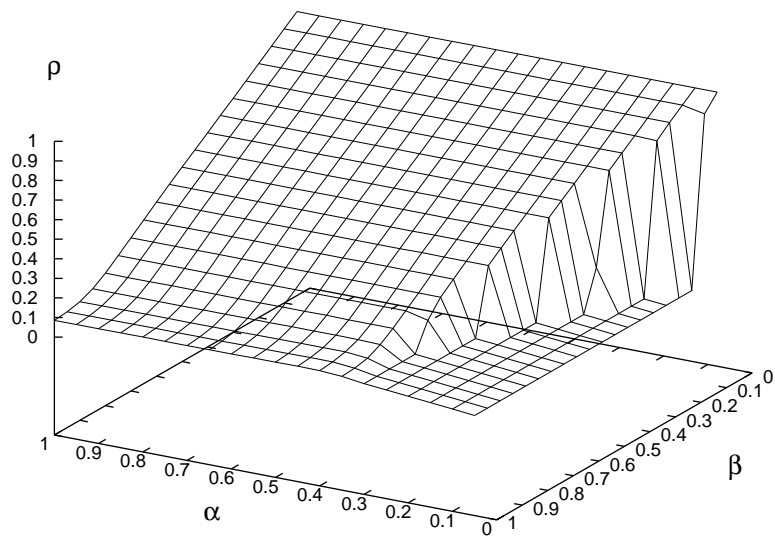
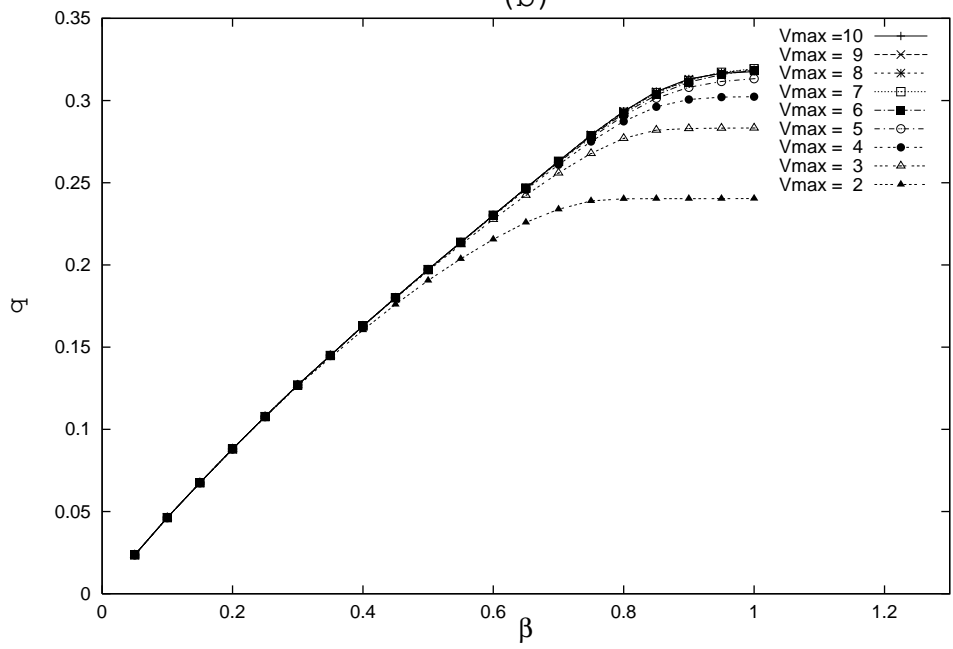
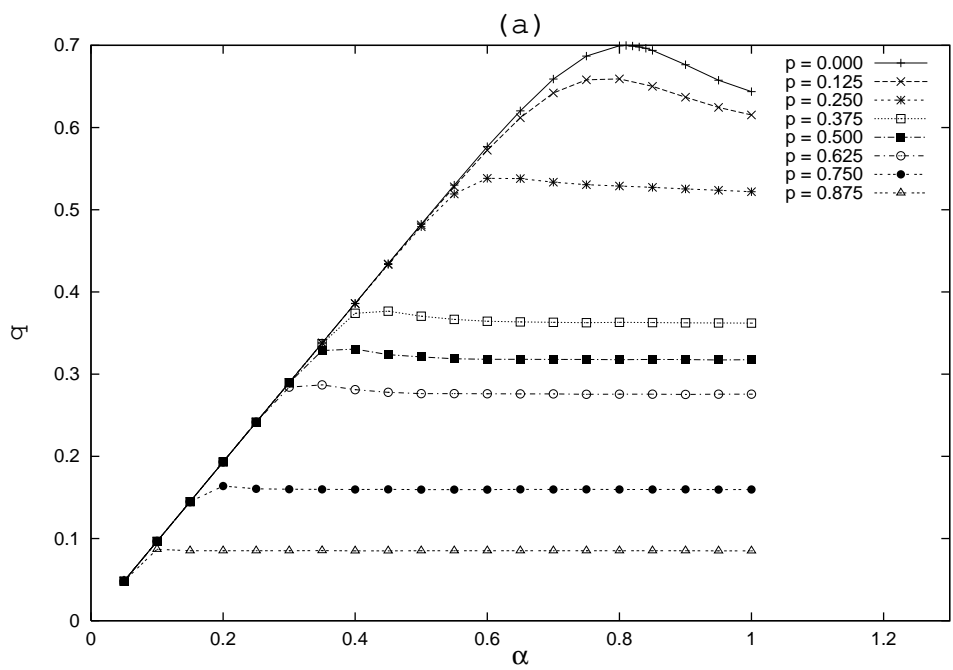
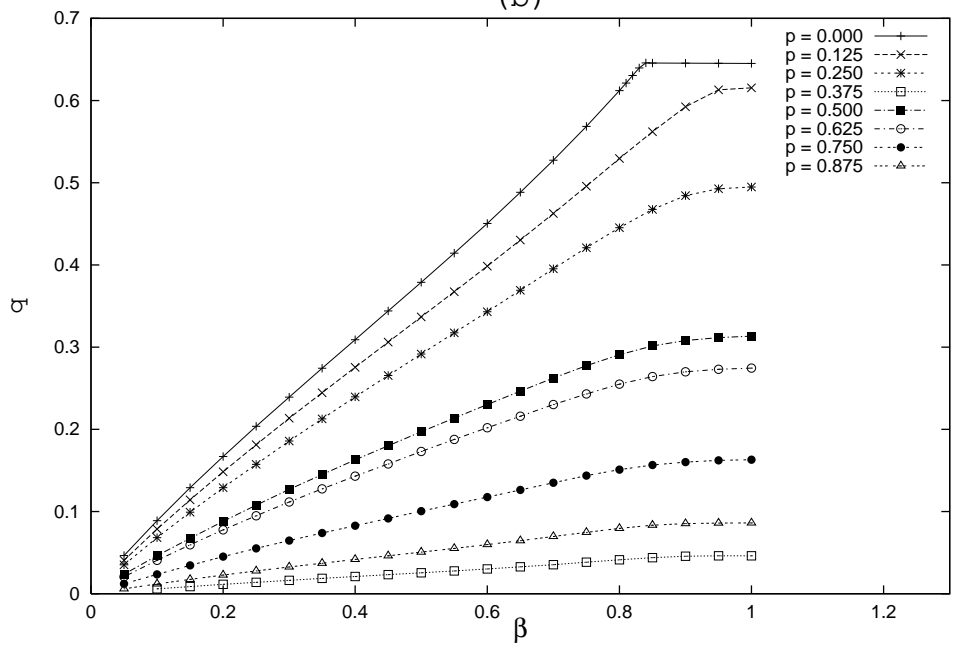
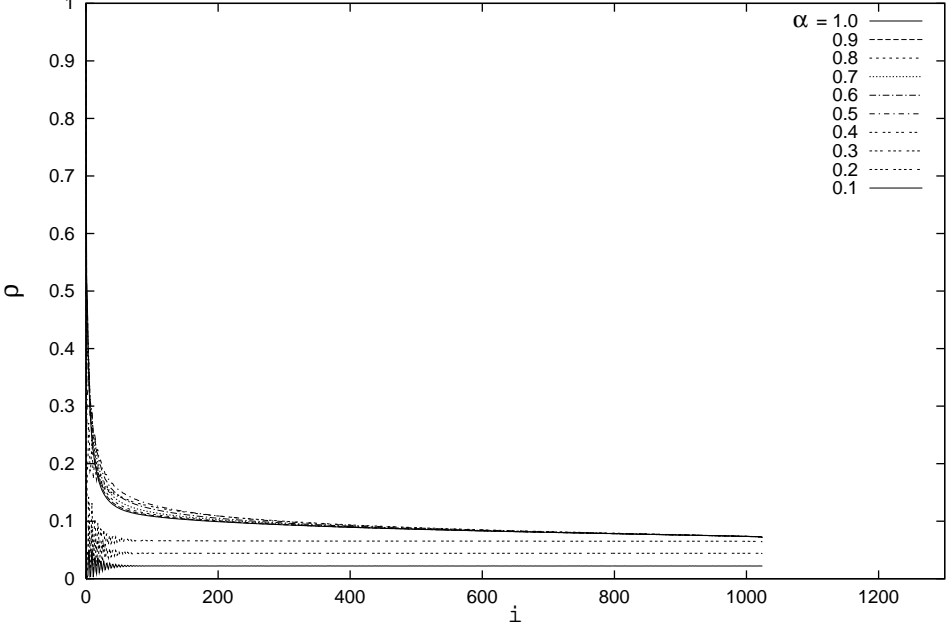
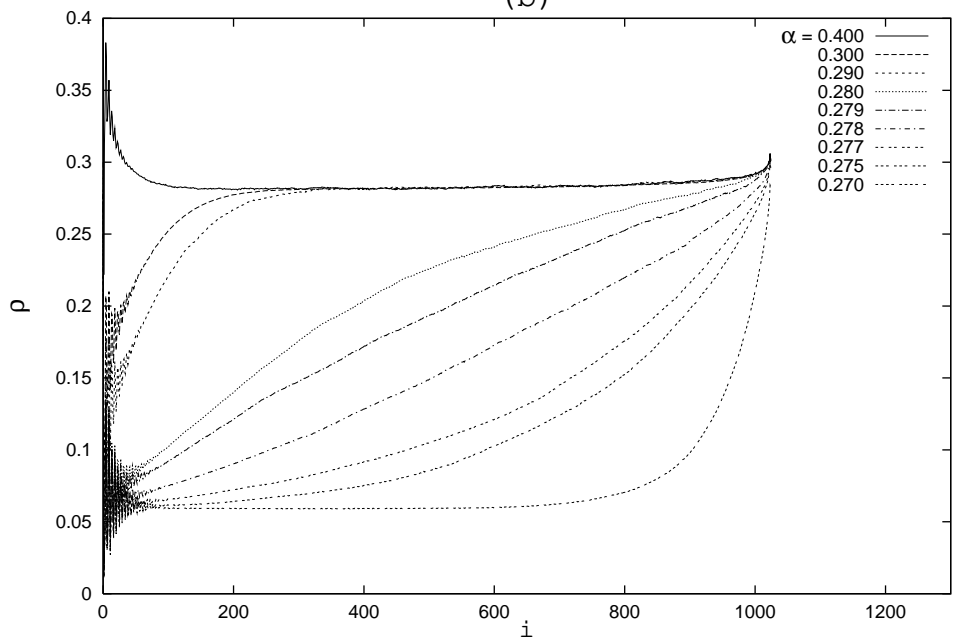
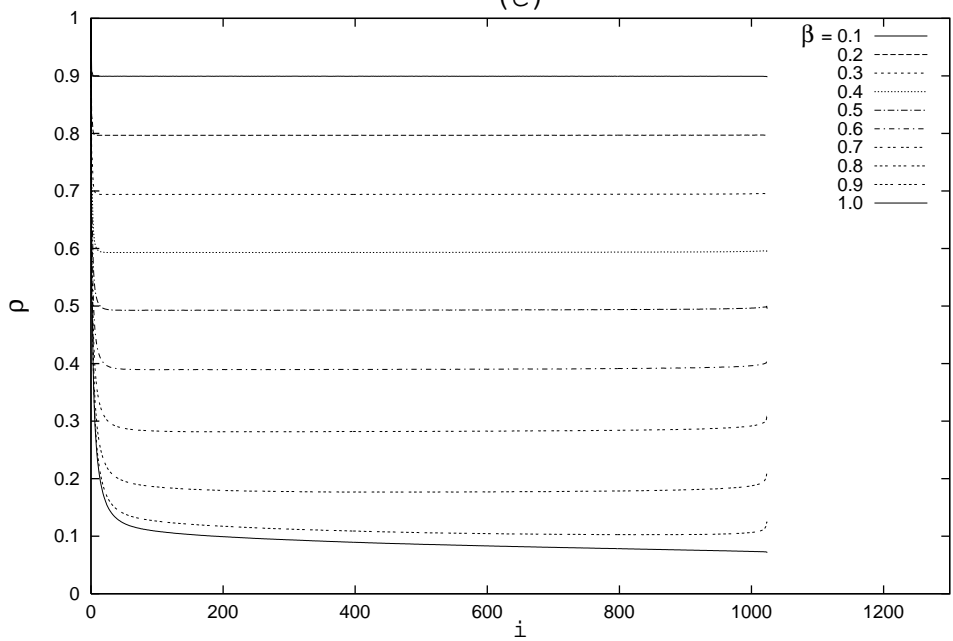
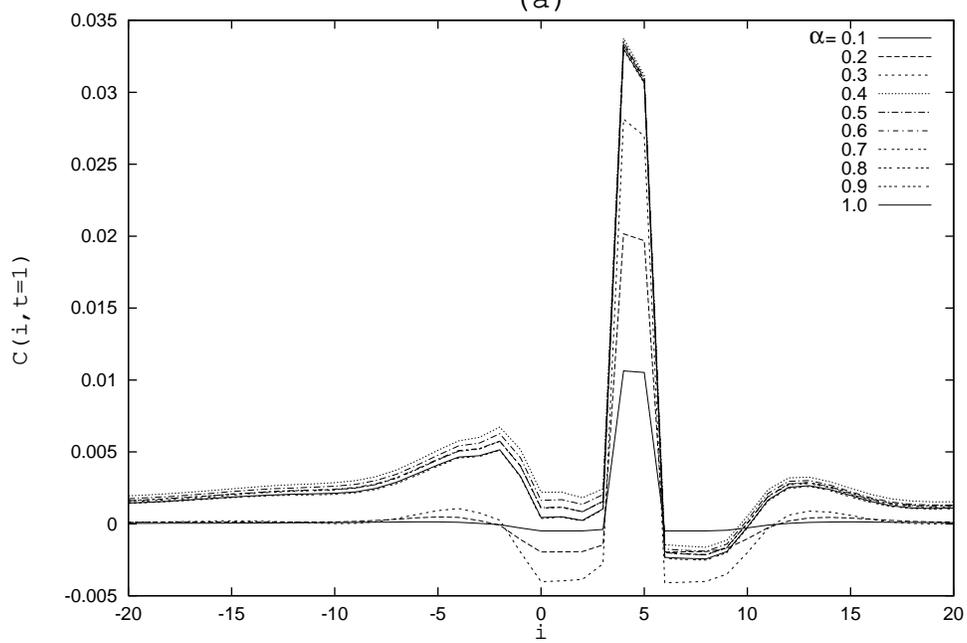
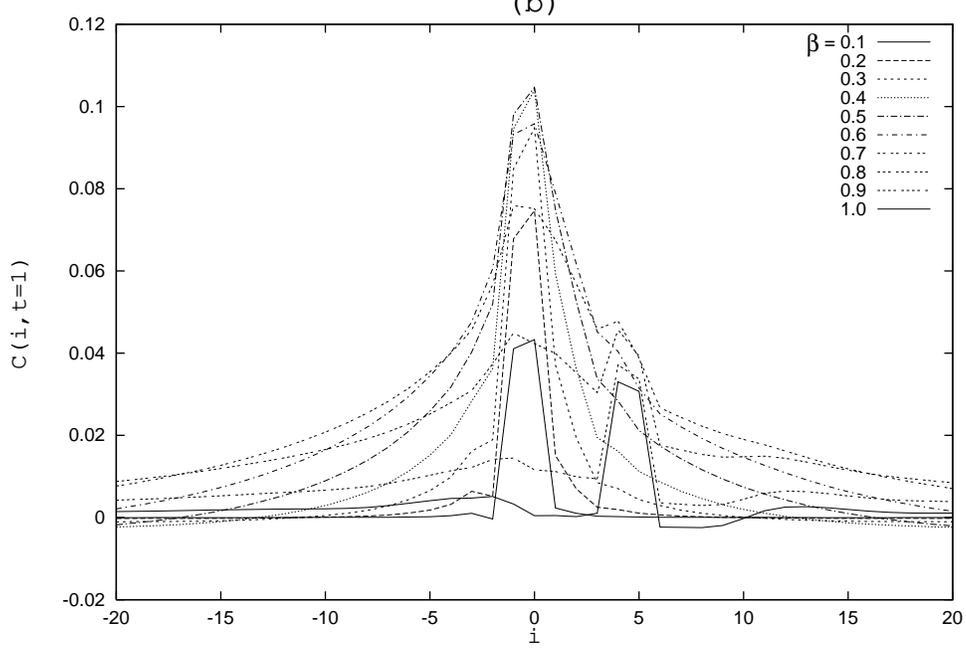
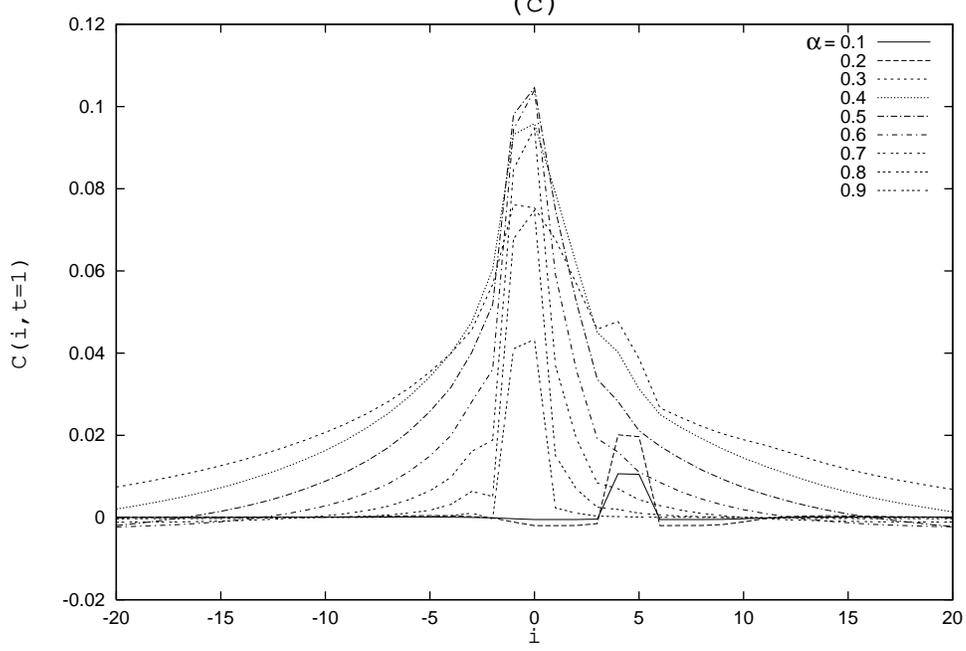
\end{document}